\documentclass[namedreferences]{SolarPhysics}
\usepackage[optionalrh,solaenum]{spr-sola-addons} % For Solar Physics
\usepackage{graphicx}
\usepackage{color}
\usepackage{url}             % For breaking URLs easily trough lines
\newcommand{\aap}{    {\it Astron. Astrophys.}}

\newcommand{\apjl}{   {\it Astrophys. J. Lett.}}

\newcommand{\pp}{     {\it Phys. Plasmas}}

\newcommand{\solphys}{{\it Solar Phys.}}

\newcommand{\ssr}{    {\it Space Sci. Rev.}}

\begin{document}
\begin{article}
\begin{opening}
\title{Modelling the Propagation of a Weak Fast-Mode MHD Shock Wave near
a 2D Magnetic Null Point Using Nonlinear Geometrical Acoustics}
\author{A.N.~\surname{Afanasyev}\sep
        A.M.~\surname{Uralov}}
\runningauthor{Afanasyev and Uralov}
\runningtitle{Fast-mode wave propagation near a magnetic null point}
\institute{Institute of Solar-Terrestrial Physics SB RAS,\\P.O. Box
291, Lermontov~St.~126A, Irkutsk 664033, Russia
email:~\url{afa@iszf.irk.ru} email:~\url{uralov@iszf.irk.ru}}
\begin{abstract}
We present the results of analytical modelling of fast-mode magnetohydrodynamic wave propagation near a 2D magnetic null point. We consider both a linear wave and a weak shock and analyse their behaviour in cold and warm plasmas. We apply the nonlinear geometrical acoustics method based on the Wentzel--Kramers--Brillouin approximation. We calculate the wave amplitude, using the ray approximation and the laws of solitary  shock wave damping. We find that a complex caustic is formed around the null point. Plasma heating is distributed in space and occurs at a caustic as well as near the null point due to substantial nonlinear damping of the shock wave. The shock wave passes through the null point even in a cold plasma. The complex shape of the wave front can be explained by the caustic pattern.
\end{abstract}

\keywords{Magnetohydrodynamics; Waves, Propagation; Waves, Shock;
Heating, Coronal}
\end{opening}

\section{Introduction}

Various eruptive processes in the solar atmosphere as well as
convective motions beneath the photosphere generate
magnetohydrodynamic (MHD) waves. Propagation of waves in an
inhomogeneous plasma can result in wave energy dissipation, which is
known to contribute into the general energy balance of the solar
corona. Behaviour of MHD waves is governed by the topology of
magnetic field and gradients of MHD quantities. Singularities of the
magnetic field topology are null points, at which the magnetic field
magnitude is equal to zero.

Dissipation of the wave energy is believed to occur most efficiently
near magnetic null points. In the solar corona, where the
plasma beta is much less than unity, a fast-mode MHD wave is subject
to significant deceleration at nulls, which therefore are the
regions of the accumulation of the wave energy. The amplitude of
the wave and gradients of all MHD quantities increase considerably
due to the Alfv\'{e}n speed decrease as well as the wave front
convergence in the vicinity of the null point, where the wave
becomes nonlinear. The wave profile steepens and the initially
linear wave transforms into a shock. Another important
effect of the increase of the gradients is the appearance of steep
spikes of the electric current density in the vicinity of the null
point, as it was demonstrated by \inlinecite{Foullon05} and modelled
numerically by \inlinecite{McLaughlin04} and
\inlinecite{Nakariakov06}. These circumstances result in rapid
conversion of the wave energy into heat.

Topological properties of magnetic null points have been studied
well (see, \textit{e.g.}, the review by \opencite{Longcope05}) and the behaviour of waves
in the vicinity of a null point has been investigated by many
authors (see the review by \opencite{McLaughlin11}).
\inlinecite{McLaughlin04} consider the incidence of a linear
fast-mode wave on a 2D null point both numerically and analytically
using the Wentzel--Kramers--Brillouin (WKB) approximation. The
authors neglect the plasma pressure against magnetic pressure and assign
the sound speed to be zero. Their analysis shows that ray paths
along which the wave travels get curved towards the null point due
to refraction. The wave envelops the null point and its energy is
converted to heat. Then \inlinecite{McLaughlin06a} consider the
wave incidence on a 2D null point formed by two magnetic dipoles. In
this case, only a portion of the wave and its energy are captured by
the null point. The rest propagates away.

The assumption of zero sound speed is a significant restriction for
the problem of interest. The solar coronal plasma has a
temperature of about $1.5 \times 10^6$ K, which gives a sound speed
of about $180$ km s$^{-1}$. There is an area around the null point,
in which the sound speed $c$ is greater than the Alfv\'{e}n speed
$V_A$. This considerably affects wave propagation. The wave is able
to pass through the null point in this case. A numerical modelling
and analytical analysis of this problem is carried out, in the
linear approximation, by \inlinecite{McLaughlin06b}. In their
numerical study, they also find conversion of the wave modes when
the wave crosses the $V_A=c$~layer.

Further, \inlinecite{McLaughlin08} consider the wave propagation
near a 3D magnetic null point under the assumption of cold plasma,
using the WKB approach. This investigation confirms capturing the
wave by the null point as well as plasma heating at that place. We
note that in the analytical studies mentioned, the authors model
only the wave front propagation and do not examine the wave
amplitude evolution.

The propagation of a nonlinear wave near a magnetic null point is
investigated by \inlinecite{McLaughlin09} for the magnetic
reconnection problem. Their numerical modelling shows a complex
pattern including the formation of a current sheet as well as the
generation of secondary shock waves. Using a similar numerical approach, \inlinecite{Gruszecki11} study
the nonlinear wave propagation to understand the triggering of the
sympathetic solar flares.

The aim of this paper is to analyse analytically nonlinear effects
of the weak fast-mode MHD shock wave propagation near a
magnetic null point and calculate the wave amplitude evolution for
the coronal heating problem. We do not study mode conversion as well
as the transformation of the null point into a current sheet by the shock
wave, since the ray method is not able to take this into
account. The paper has the following outline. We consider the linear
wave case in Section~\ref{S-linear wave}, then the case of a shock
wave in Section~\ref{S-shock wave} and discuss the results obtained.
Section~\ref{S-conclusion} contains concluding remarks. In the
Appendices we give bulky mathematical details of our investigation.

\section{Linear wave}
\label{S-linear wave}

In this study we explore a two-dimensional magnetic null point,
which is located at the point of origin and specified in Cartesian
coordinates $(x,y,z)$ as $\mathbf{B}=(x,0,-z)$ (see, \textit{e.g.},
\opencite{Parnell96}). In this model, the magnetic field magnitude
grows linearly with the distance from the null point. The plasma
density $\rho$ and temperature $T$ are assumed to be constant, so
the sound speed $c=\sqrt{2 \gamma R_{gas} T/\widetilde m M_H}$ is
also constant and the Alfv\'{e}n speed $V_A=B/\sqrt{4 \pi \rho}$
distribution is governed only by that of the magnetic field. Here
$M_H$ is the molar mass of hydrogen, $\widetilde{m}$ the average
atomic weight of an ion, $\gamma$ the adiabatic index, and $R_{gas}$
the gas constant. The Alfv\'{e}n speed vanishes completely at the
null point and grows to infinity away from it.

\subsection{Cold plasma}
\label{S-linear wave-cold plasma}

At first we consider the case of a cold plasma, with the sound speed
$c$ being zero overall. Let a linear plane fast-mode MHD wave
propagate towards the magnetic null point. To calculate its
propagation, we apply the geometrical acoustics method (WKB
approach). In this method, a solution is found in the form of
$A\left( \mathbf{r},t\right) e^{i\Psi \left( \mathbf{r},t\right) }$
where $A\left( \mathbf{r},t\right)$ is the wave amplitude, and $\Psi
\left( \mathbf{r},t\right)$ is the eikonal, depending on both
coordinates $\mathbf{r}=(x, z)$ and time $t$. By substituting this
representation for wave perturbations into the set of linearised
equations of ideal magnetohydrodynamics, one can obtain a
Hamilton--Jacobi partial differential equation for the eikonal of
fast-mode MHD waves (see, \textit{e.g.}, \opencite{Uralova94}):
$$\frac{\partial \Psi }{\partial t}+a\left\vert \mathrm{grad}\Psi
\right\vert =0,$$ where $a=1/2 \left( \sqrt{c^2+V_A^2+2 c V_A \cos
\alpha} + \sqrt{c^2+V_A^2-2 c V_A \cos \alpha} \right)$ is the
fast-mode velocity in plasma, $\alpha$ the angle between the
direction of the wave propagation and the magnetic field vector.
Solving the equation with the method of characteristics gives the
set of ray equations, which are a set of ordinary differential
equations. In Appendix \ref{S-appendix1} we provide this set of
equations (see Equations~(\ref{E-ray_eq_system})).
\inlinecite{McLaughlin04} succeeded in finding an exact analytical
solution of the set for the magnetic field model under
consideration:
\begin{equation}
x=e^{- \frac {z_0} {r_0} t } \left(  x_0 \cos \frac {x_0} {r_0} t
+ z_0 \sin \frac {x_0} {r_0} t \right),\,
z=e^{- \frac {z_0} {r_0} t } \left(  z_0 \cos \frac {x_0} {r_0} t
- x_0 \sin \frac {x_0} {r_0} t \right),
\label{E-xz-solutions}
\end{equation}
where $r_0=\sqrt{x_0^2+z_0^2}$,  and $x_0$ and $z_0$ are the
initial values of a ray path.

\begin{figure} % {1}
  \centerline{\includegraphics[width=0.5\textwidth]{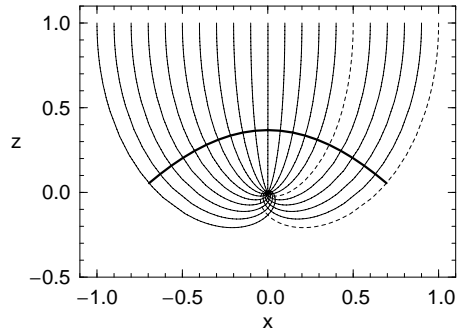} }
\caption{The propagation of a linear wave near a null point in a
cold plasma. The thin lines represent the rays and the thick one
marks the wave front at moment $t=1$. At the initial moment, the
plane wave front was located at $z=1$ and $-1 \leq x \leq 1$.
The axes are shown in the normalised units.}
  \label{F-prop_pattern_lin_cold}
\end{figure}

The null point attracts the wave, so all the ray paths tend to the
origin where the null point is located. Figure
\ref{F-prop_pattern_lin_cold} shows the ray pattern. In both the
figures and calculations throughout the paper, we use
normalised units. The normalising values are the characteristic spatial scale of the problem, and the magnetic field strength and the Alfv\'{e}n speed at the distance of the characteristic spatial scale from the null point. The normalising time value is then set as the ratio of the characteristic spatial scale to the characteristic Alfv\'{e}n speed.

However, it is essential not only to find the wave geometry, but
also to calculate the wave intensity. Geometrical acoustics allows
the wave amplitude variation to be calculated.

In linear geometrical acoustics, the energy flux of a disturbance
travelling at group velocity $\mathbf{q}$ in a stationary medium is
directed along the rays, and its magnitude is conserved within a ray
tube \cite{Blokhintsev81}, $\mathrm{div}(\Delta \varepsilon
\mathbf{q}) = 0$, where $\Delta \varepsilon =\rho \overline{\left(
u^{2}+v^{2}\right) }$ is the average density of the disturbance
energy, $\rho$ the undisturbed plasma density, and $u,v$ the plasma
velocity components along the normal $\mathbf{k}$ to the wave front
and across it, respectively. Taking into account the relation
between the plasma velocity components \cite{Kulikovsky05}, $\mu =
v/\,u= \left( 1- c^2/\, a^2 \right) \cot \alpha$, it is possible to
relate the variation of the wave amplitude to the cross section $dS$
of the ray tube formed by a bundle of rays:
\begin{equation}
 dSq\rho u^{2}\left( 1+\mu ^{2}\right) =const.
 \label{E-dS_wave_amp_variation}
\end{equation}

We calculate the cross section, using the Jacobian of the
transformation from Cartesian coordinates to ray ones
\cite{KravtsovOrlov90}. The volume element $dW$ of the ray tube is
expressed in terms of the ray coordinates $(\eta , t)$ as
$$dW=dxdz= D\left(
t\right) d\eta dt, \quad D\left(
t\right)= \left \vert \begin{array}{l}
{\partial x/\partial \eta} \quad {\partial x/\partial t} \\
{\partial z/\partial \eta} \quad {\partial z/\partial t}
\end{array} \right \vert ,$$
where $D(t)$ is the Jacobian. Then for the cross section of the ray tube, $dS$, we have
$$dS=\frac{dW}{d\sigma }=\frac{D\left( t\right)
}{q}d\eta,$$
where $\sigma$ is the ray length. Let the ray coordinate $\eta$ be
the initial value $x_0$. Using Equations (\ref{E-xz-solutions}) and
(\ref{E-dS_wave_amp_variation}) we obtain the expressions for the
wave amplitude $A$ along the ray with the initial values $x_0$ and
$z_0$ ($A_0$ is the initial amplitude):
\begin{equation}
A \left(
t\right)/\, A_0 = \left \vert \frac {x_0 \cos 2 \frac {x_0} {r_0} t + z_0 \sin 2 \frac {x_0} {r_0} t} {x_0} \right \vert \sqrt{\frac {r_0} {z_0 t +r_0}} \;\; e^{\frac {z_0} {r_0} t},
\label{E-wave_amp_expr_u}
\end{equation}
if we assume the wave amplitude $A$ to be the component $u$ of the plasma velocity vector along the normal to the wave front, or
\begin{equation}
A \left(
t\right)/\, A_0= \sqrt{\frac {r_0} {z_0 t +r_0}} \;\; e^{\frac {z_0} {r_0} t},
\label{E-wave_amp_expr_magnitude}
\end{equation}
if we assume $A$ to be the magnitude of the plasma velocity vector,
$A=\sqrt{u^2+v^2}$.

\begin{figure} % {2a+b}
  \centerline{  \includegraphics[width=0.5\textwidth,clip=]{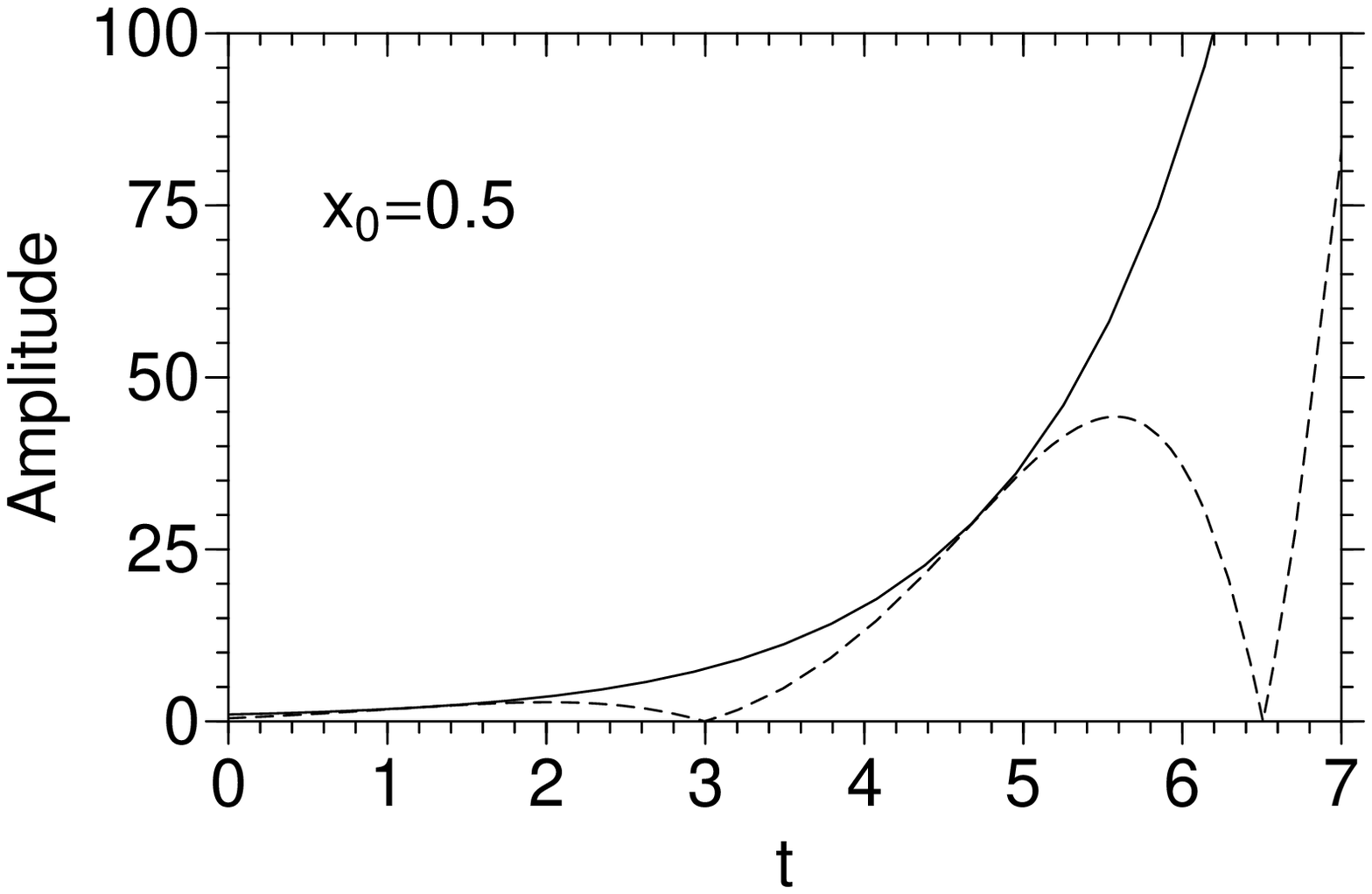}
                \includegraphics[width=0.5\textwidth,clip=]{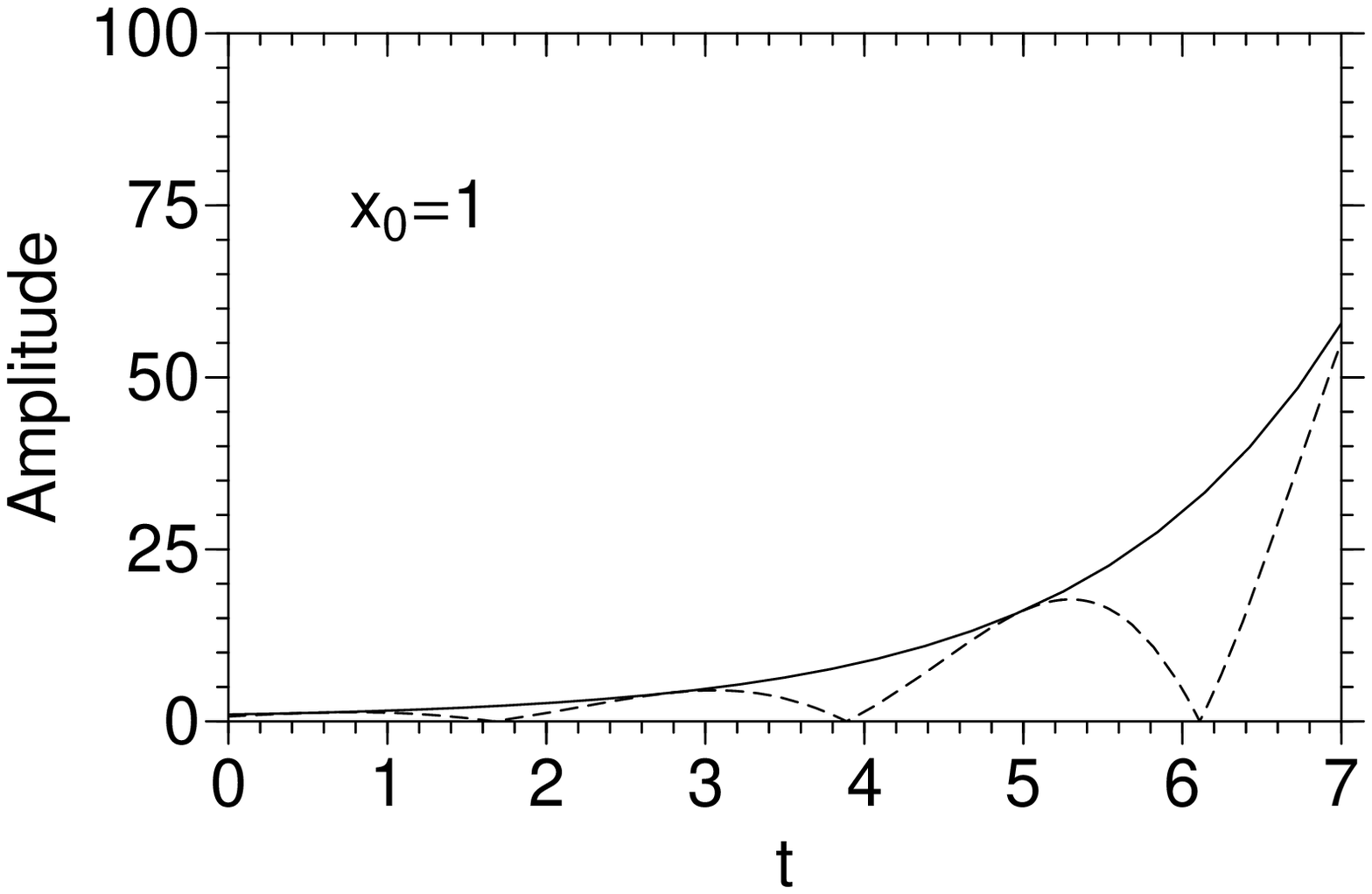}
             }
    \vspace{-0.11\textwidth}
          \centerline{\large
      \hspace{0.1 \textwidth}  {a}
      \hspace{0.47 \textwidth} {b}
         \hfill}
   \vspace{0.075\textwidth}
\caption{Evolution of the linear wave amplitude along the rays with
$x_0=0.5$ (a) and $x_0=1$ (b) (marked as the dashed lines in Figure
\ref{F-prop_pattern_lin_cold}) in the cold plasma case. The dashed
lines represent the plasma velocity component $u$ along the normal
to the wave front and the solid ones show the magnitude of the
plasma velocity. The horizontal axes are shown in the
normalised time units, the vertical axes are shown in the units of
the initial amplitude.}
  \label{F-amplitude_lin_cold}
\end{figure}

Figure \ref{F-amplitude_lin_cold} shows time evolution of the wave
amplitude along two different rays. The amplitude plots in
the paper are shown in the units of the initial amplitude. The
plasma velocity component $u$ oscillates and drops to zero from time
to time, while the magnitude of the plasma velocity grows
exponentially. Such behaviour of the component $u$ is due to
changing the direction of the wave propagation. The wave front
element travels around the null point and sometimes the wave front
normal $\mathbf{k}$ becomes collinear to the magnetic field
$\mathbf{B}$, \textit{i.e.} $\alpha = 0$ or $\pi$. At these points,
the component $u$ disappears but the component $v$ becomes equal to
the magnitude of the plasma velocity vector at that moment. There is
also the opposite case: when the $\alpha=\pm \, \pi/2$ the component
$v$ drops to zero. So the solid line marking the magnitude of the
plasma velocity is an envelope line. Note that in Figure
\ref{F-amplitude_lin_cold} we plot a relative variation of the
amplitude and adjust the magnitude of the plasma velocity vector to
be unity at the initial moment. It means that for the component $u$
we actually use Equation~(\ref{E-wave_amp_expr_u}) divided by
$r_0/|x_0|$.

\subsection{Warm plasma}
\label{S-linear wave-warm plasma}

Now we consider the case of a warm plasma with non-zero sound speed.
Let points where the Alfv\'{e}n speed is equal to the sound speed be
named the $V_A=c$ layer (see, \textit{e.g.},
\opencite{McLaughlin06b}). In the problem of interest, the
case of a warm plasma corresponds to the situation when the wave
propagates near the $V_A=c$ layer, while in the case of a cold
plasma the wave propagation is considered far from it. The ray
equations become more complex and we have to integrate them
numerically. To calculate the Jacobian and find the wave amplitude,
we use a method based on numerical integration of the so-called
adjoint set of equations. Solving the adjoint set allows the
quantities $\partial x/\partial \eta$ and $\partial z/\partial \eta$
to be calculated. For more details, we refer the reader to the paper
by \inlinecite{Afanasyev11}. Appendix \ref{S-appendix1} contains the
set of ray equations (see
Equations~(\ref{E-ray_eq_system})) as well as the adjoint set
(see Equations~(\ref{E-adj_system})), which we solve to
model the wave propagation.

\begin{figure} % {3a+b}
  \centerline{\includegraphics[width=1\textwidth,clip=]{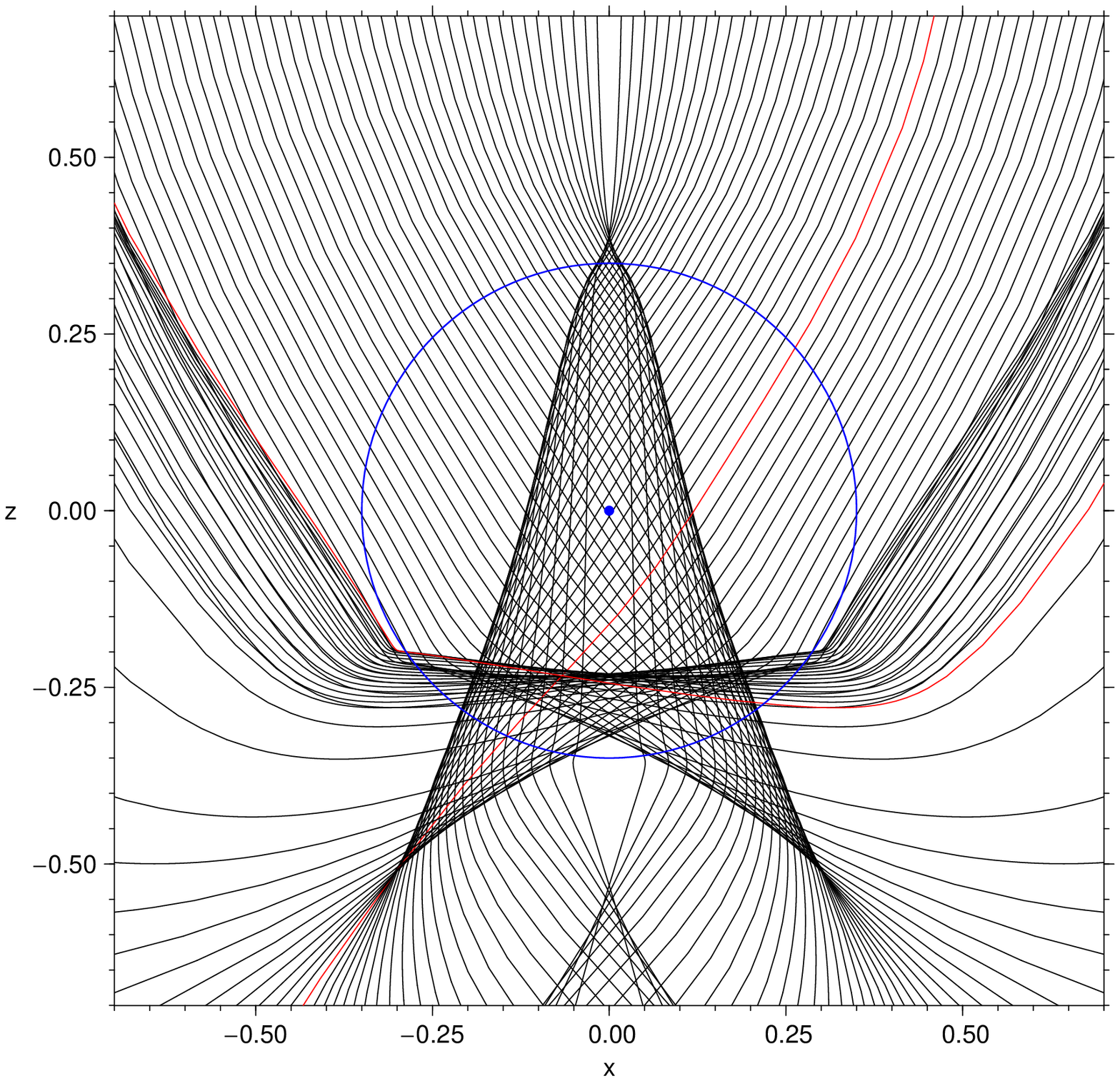}
             }
  \centerline{\includegraphics[width=0.5\textwidth,clip=]{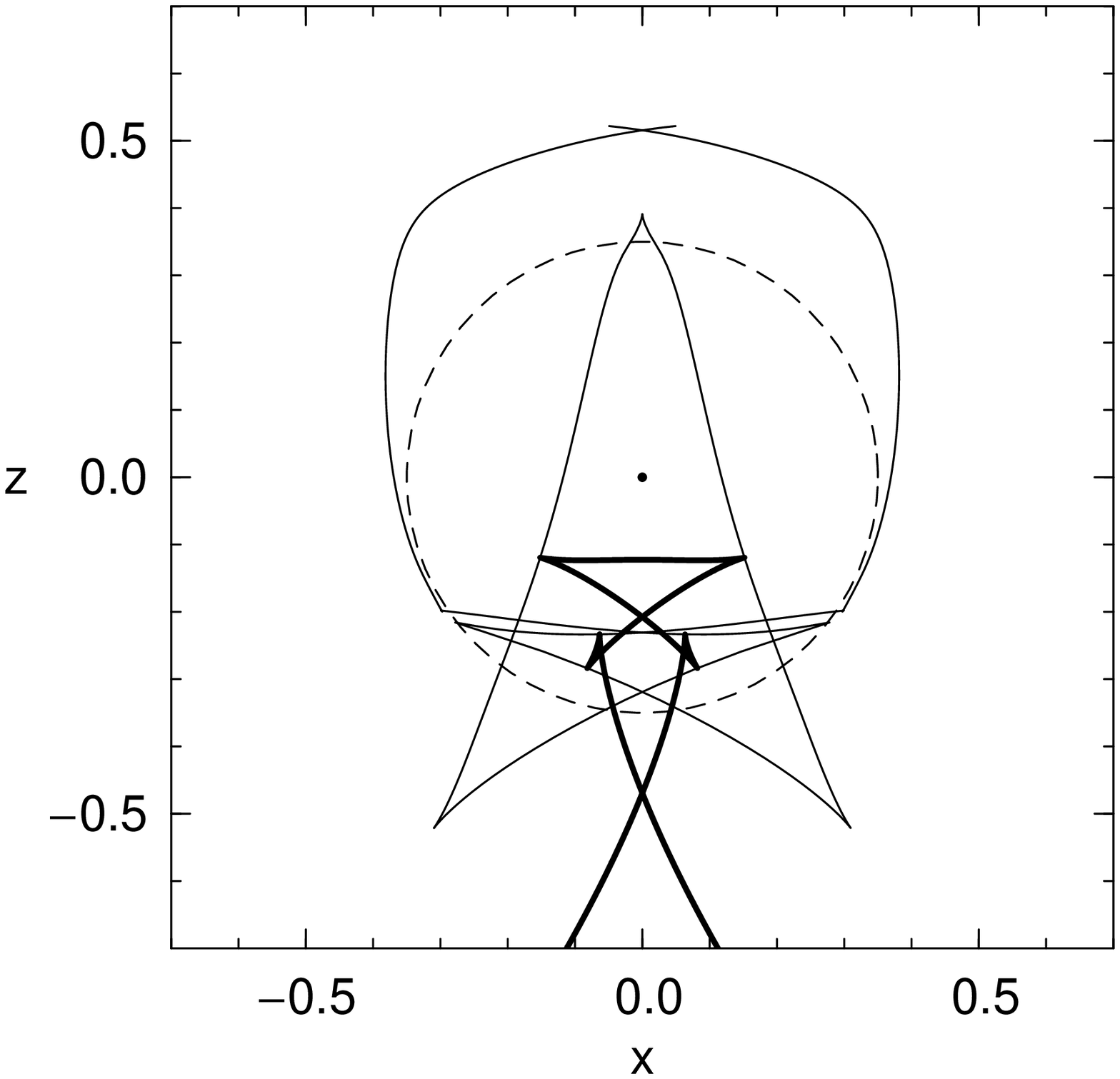}
             }
\caption{The propagation of a linear wave near a null point in a
warm plasma: the ray pattern (top), and the wave front and the
caustic (bottom). In the top panel, blue denotes the location of the
$V_A=c$~layer and the null point, black and red -- the rays. In the
bottom panel, the solid lines denote the wave front at moment
$t=2.4$ (thick) and the caustic (thin), and  the dashed line denotes
the $V_A=c$~layer. At the initial moment, the plane wave front was
located at $z_0=1$ and $-2 \leq x \leq 2$. The axes are
shown in the normalised units.}
  \label{F-prop_pattern_lin_warm}
\end{figure}

Like \inlinecite{McLaughlin06b} we find that the wave passes through
the null point. Figure \ref{F-prop_pattern_lin_warm} shows the wave
propagation in this case. We study behaviour of the wave front, which
was located at $z = 1$ and $-2 \leq x \leq 2$ at the initial moment,
however we plot the rays outgoing from a central part of the wave
front for clearing the ray pattern. Unlike the cold plasma case, ray
paths begin to intersect near the top boundary of the $V_A=c$ layer.
This results in the considerable distortion of the wave front as
well as the subsequent formation of a caustic.

\begin{figure} % {4a+b}
  \centerline{  \includegraphics[width=0.5\textwidth,clip=]{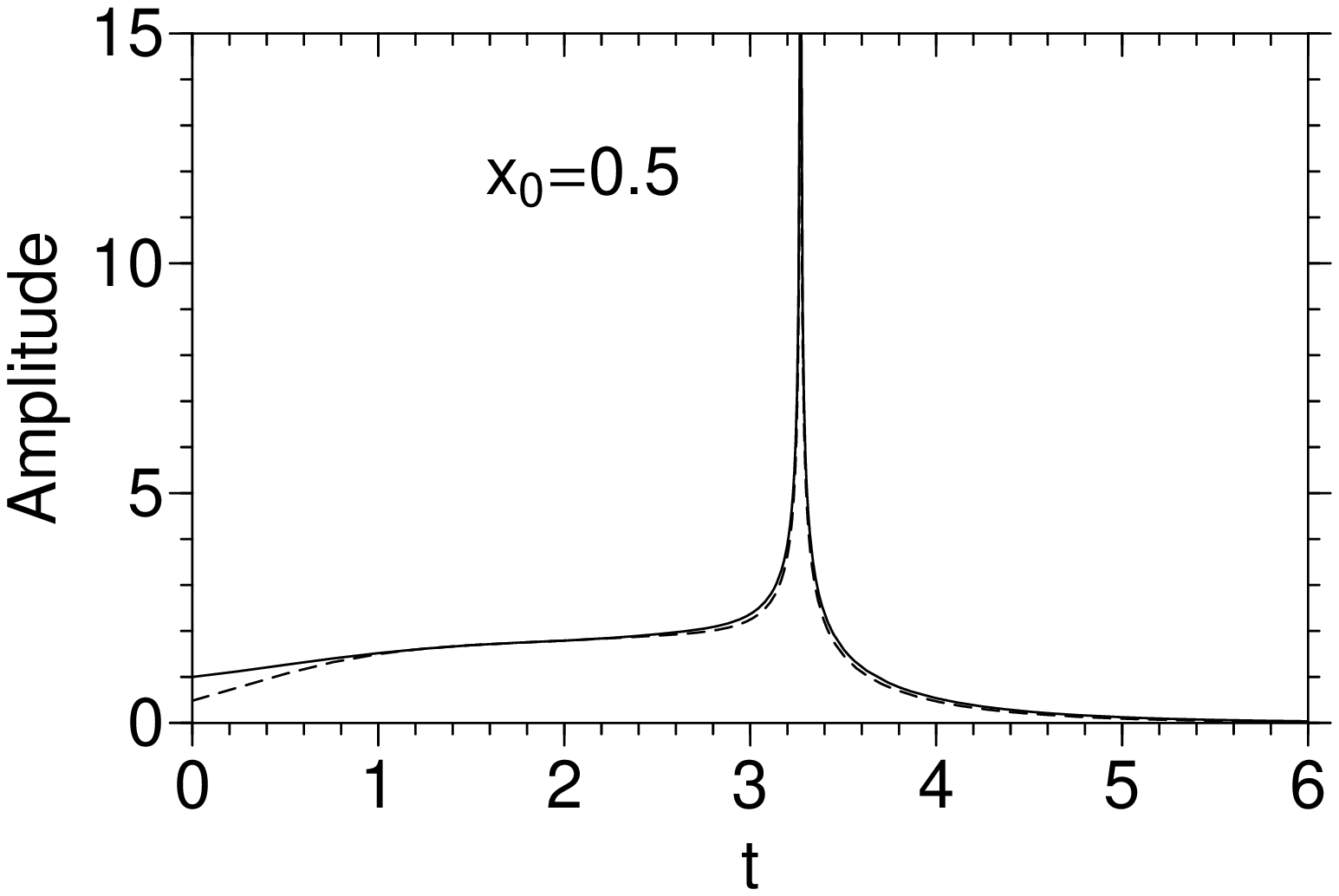}
                \includegraphics[width=0.5\textwidth,clip=]{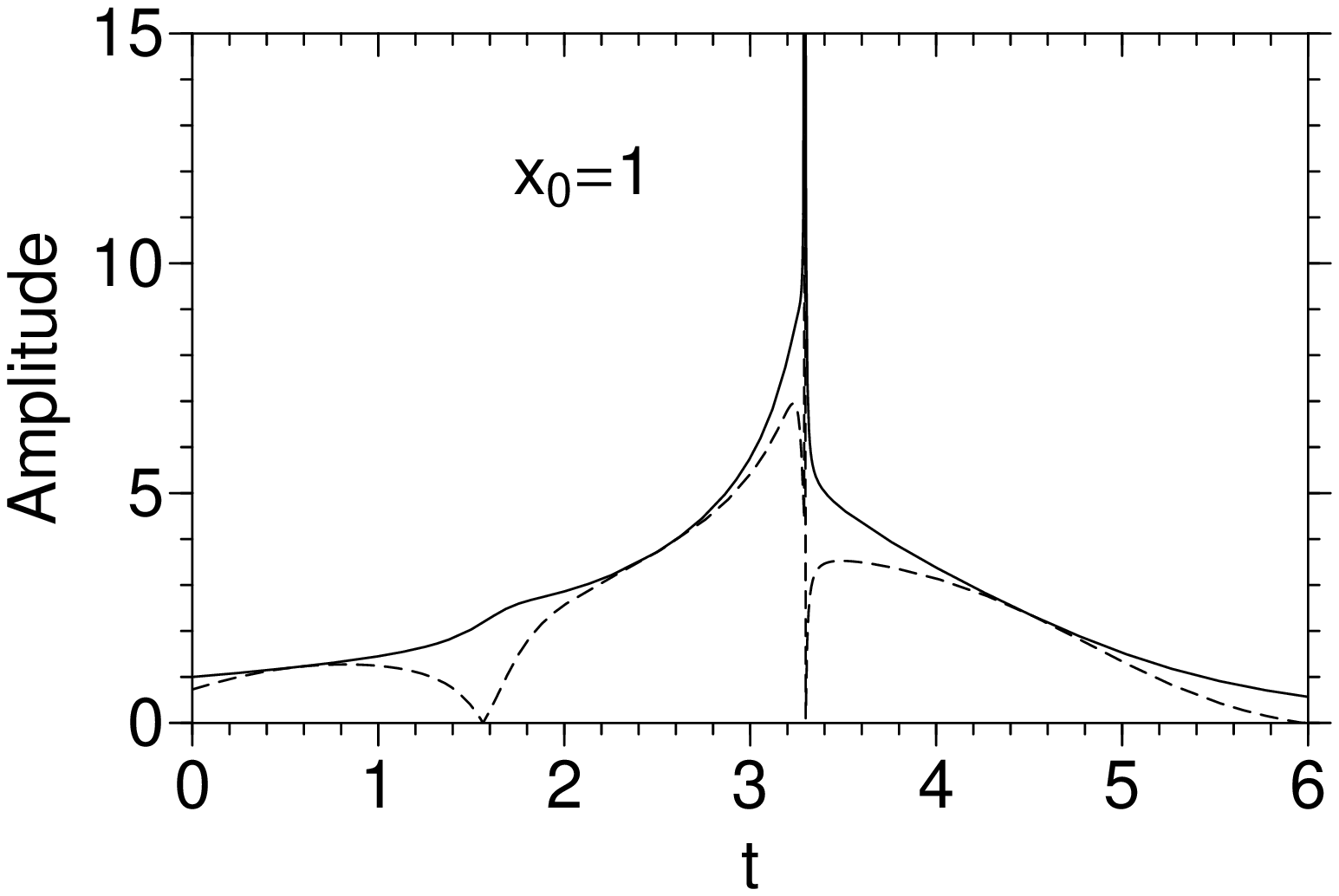}
             }
    \vspace{-0.28\textwidth}
          \centerline{\large
      \hspace{0.41 \textwidth} {a}
      \hspace{0.47 \textwidth} {b}
         \hfill}
   \vspace{0.245\textwidth}
\caption{Evolution of the linear wave amplitude along the rays with
$x_0=0.5$ (a) and $x_0=1$ (b) (red lines in Figure
\ref{F-prop_pattern_lin_warm}) in the warm plasma case. The dashed
lines represent the plasma velocity component $u$ along the normal
to the wave front and the solid ones show the magnitude of the
plasma velocity. The horizontal axes are shown in the
normalised time units, the vertical axes are shown in the units of
the initial amplitude.}
  \label{F-amplitude_lin_warm}
\end{figure}

By definition, a caustic is an envelope of a family of rays,
\textit{i.e.} rays approach each other at points of the caustic
\cite{KravtsovOrlov90}, and therefore a caustic can be identified in
the ray pattern as a ray crowding. The Jacobian and the ray-tube
cross sections drop to zero at the caustic points and the wave
amplitude grows indefinitely. Figure~\ref{F-amplitude_lin_warm}
shows evolution of the wave amplitude along two rays outlined by red
in Figure \ref{F-prop_pattern_lin_warm}. Marking points of the ray
paths where the amplitude tends to infinity for the first time
(since this can occur twice along some rays during the calculation
time in our modelling) we can obtain the caustic line. The bottom
panel in Figure \ref{F-prop_pattern_lin_warm} shows the inner part
of the caustic, associated with a central part of the wave. The rest
of the caustic encompasses the null point, bending away from it. The
outer parts of the caustic form if the initial front is rather
extensive. Besides, if the calculation time is also rather long it
is possible to identify secondary (repeated) caustics. For instance,
one of them can be seen in the top panel of Figure
\ref{F-prop_pattern_lin_warm} as a cusp at the lower part of the
plot. Note that in the linear cold plasma case there are no caustics
in the neighbourhood of the null point.

The singular behaviour of the wave amplitude indicates that the ray
approximation does not work at the caustic. Indeed, more accurate
considerations show that the amplitude grows but it always remains
finite \cite{KravtsovOrlov90}. Nevertheless, caustics give important
information on the amplitude distribution along the wave front. At
every moment, the wave amplitude increases significantly at a few
points (the caustic does not ``fire up'' simultaneously throughout
its extent). Caustics in the ray pattern represent the places of the
most efficient plasma heating associated with the wave. The energy
of the wave accumulates in the neighbourhood of the null point, but
this accumulation is distributed in space, unlike the cold plasma
case. Our calculations show that the size of the $V_A=c$ layer may
be chosen as the characteristic spatial scale of the heated
region around the null point.

Analysing the caustic pattern we can also readily explain the
complicated surface of the wave front mentioned in
\inlinecite{McLaughlin06b}. A break of the wave front occurs at the
caustic (see Figure \ref{F-prop_pattern_lin_warm}), so ``small
triangular shapes'' of the wave front arise at the caustic cusps.
Vertices of these triangles travel along the caustic lines as the
wave front propagates.

\section{Shock wave}
\label{S-shock wave}

As seen in Section \ref{S-linear wave-cold plasma}, the wave
amplitude grows exponentially as the wave approaches the null point.
This results in the nonlinear steepening of the wave front and
therefore a shock wave formation. So we should take into account
nonlinear properties of shock waves and investigate how they affect
the wave propagation.

\begin{figure} % {5a+b}
  \centerline{  \includegraphics[width=0.495\textwidth,clip=]{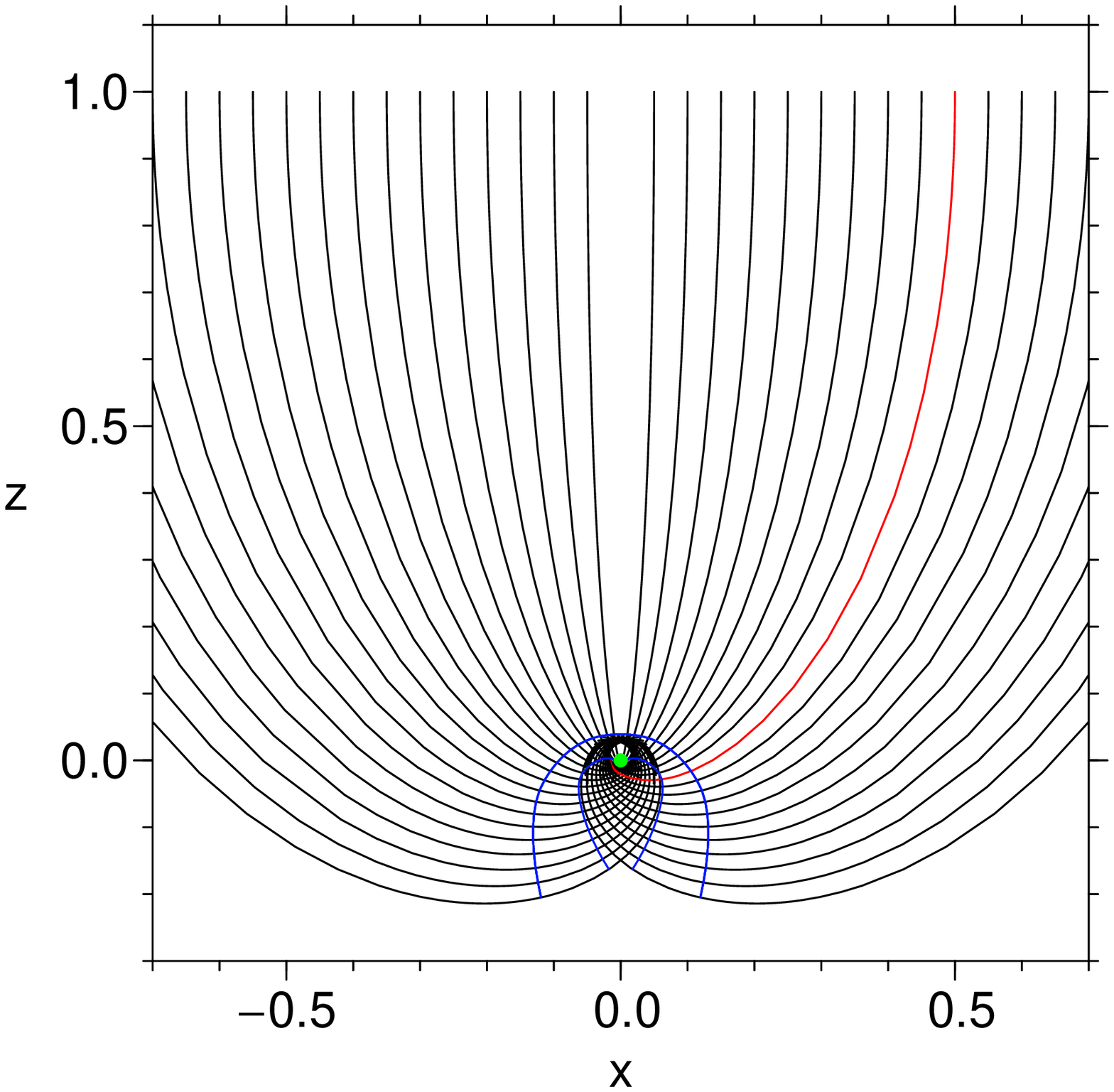}
                \includegraphics[width=0.505\textwidth,clip=]{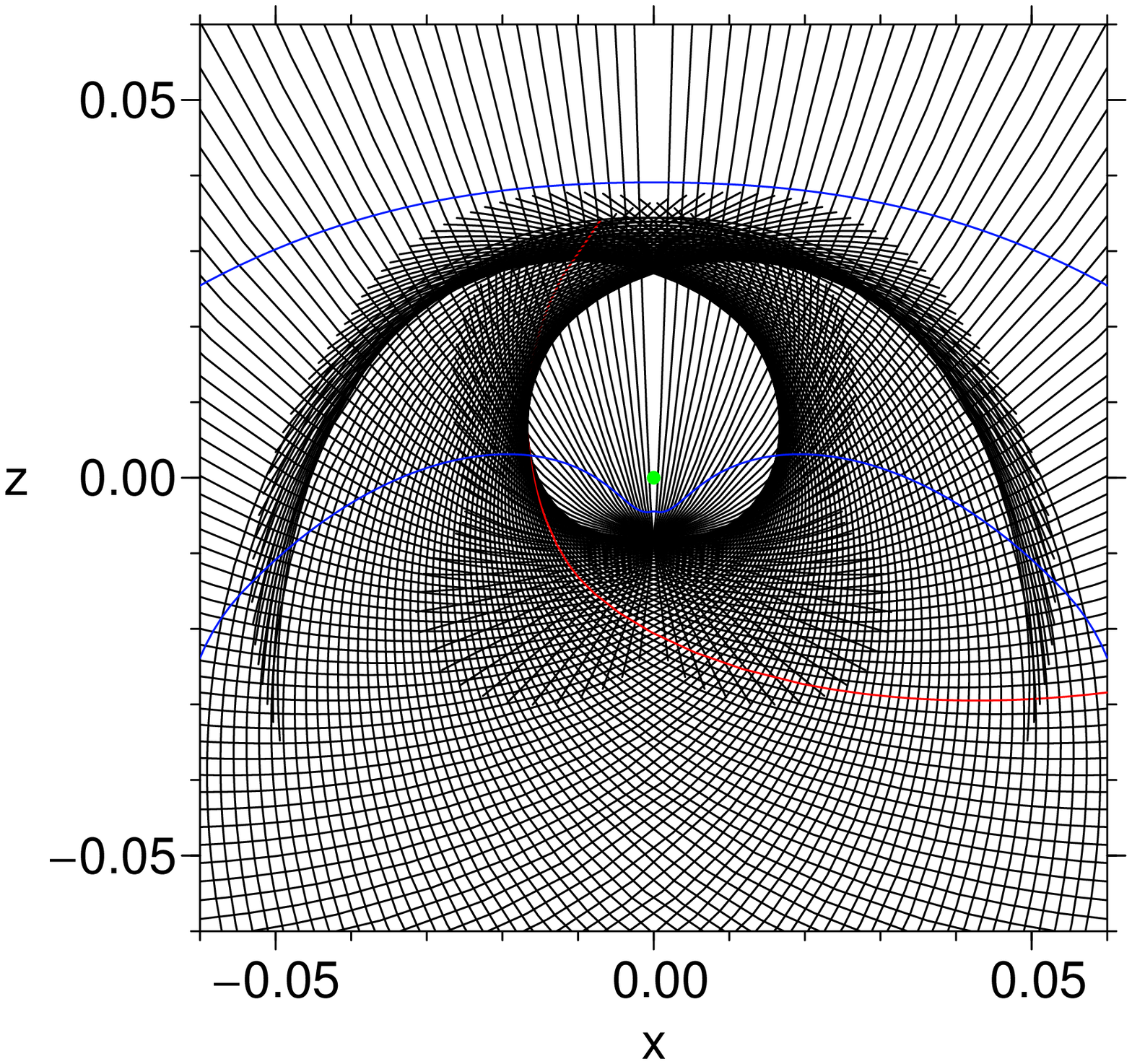}
             }
    \vspace{-0.051 \textwidth}
          \centerline{\large
      \hspace{0.01 \textwidth} {a}
      \hspace{0.49 \textwidth} {b}
         \hfill}
   \vspace{0.016 \textwidth}
\caption{The propagation of a shock wave near a null point in a cold
plasma. Blue colour denotes the positions of the wave front at
moments $t=2.5$ and $t=3.0$, green -- the null point, black and red
-- the rays. At the initial moment, the plane wave front was located
at $z_0=1$ and $-1 \leq x \leq1$. Panel b shows a zoom on the null
point vicinity. The axes are shown in the normalised
units.}
  \label{F-prop_pattern_shock_cold}
\end{figure}

Firstly, a nonlinear plane disturbance in an ideal
homogeneous medium propagates at speed that exceeds the fast-mode
velocity and depends on the disturbance amplitude. In a nonlinear
acoustics approximation, a fast-mode wave element with plasma
velocity component $u$ normal to the front moves at a speed of
$a+\kappa u$, where $\kappa = \left( 1/a\right) \left(d\left( \rho
a\right) /d\rho \right) $ is the numerical coefficient depending
both on the plasma beta and the angle $\alpha$ between the wave
front normal and the magnetic field (see Appendix
\ref{S-appendix1}). The fact that each element of the wave travels
at its own speed causes the wave profile deformation and the
appearance of a shock wave. In turn, a weak shock moves at a speed
of $a+\kappa U_{sh}/2$, where $U_{sh}$ is the jump of the plasma
velocity component $u$ in the shock.

Secondly, the nonlinear wave amplitude undergoes additional damping
associated with the energy dissipation in the discontinuity. On the
one hand, the shock wave amplitude increases due to the convergence
of rays as well as the Alfv\'{e}n speed decrease. On the other hand,
the amplitude decreases due to the nonlinear damping.

The above-mentioned basic properties of shock waves are taken into
account in the method of nonlinear geometrical acoustics. It allows
one to calculate the propagation of disturbances with small but
finite amplitudes (\textit{i.e.} weak shock waves) through an
inhomogeneous medium. To familiarise the readers with details of the
method used, we refer them to the papers by \inlinecite{Afanasyev11}
and \inlinecite{Uralova94}. Besides, in Appendix \ref{S-appendix1}
we give the set of nonlinear ray equations (see
Equations~(\ref{E-ray_eq_system})), which we solve to model the
weak shock wave propagation, and in Appendix \ref{S-appendix2} we
present derivation of the nonlinear ray equations.

\begin{figure} % {6}
  \centerline{\includegraphics[width=0.5\textwidth]{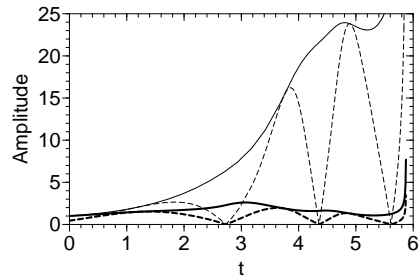} }
\caption{Evolution of the shock wave amplitude along the ray with
$x_0=0.5$ in the cold plasma case. The thick lines correspond to the
amplitude of the shock wave and the thin ones correspond to the
amplitude variation due to only the convergence of rays and the
$V_A$~decrease (without dissipation in the discontinuity). The
dashed lines denote the jump $U_{sh}$ of the plasma velocity
component $u$ in the shock wave, the solid ones -- that of the
plasma velocity magnitude. The horizontal axis is shown in
the normalised time units, the vertical axis is shown in the units
of the initial amplitude.}
  \label{F-amplitude_shock_cold}
\end{figure}

Figure \ref{F-prop_pattern_shock_cold} shows the propagation of a
weak shock wave in a cold plasma. Like the linear cold
plasma case, the wave is captured by the null point and wraps around it.
However, the propagation speed at the null point is non-zero owing
to nonlinearity and the shock wave passes through the null point
(cf. \opencite{McLaughlin09}) like a linear wave in a
plasma with the small sound speed. Therefore, in this case a caustic
also arises in the small neighbourhood of the null point. This can
be seen in Figure \ref{F-prop_pattern_shock_cold}b, which shows the
propagation pattern in a small area around the null point. The
caustic is seen as the sharp border of a ray crowding. The amplitude
of the wave tends to infinity at the caustic. This indicates again
the places of significant plasma heating.

The wave front elements being close to the null point have
higher amplitudes and therefore they travel faster. This results in
forming a bulge in the wave front, which is visible in
Figure \ref{F-prop_pattern_shock_cold}b underneath the null
point.

Figure \ref{F-amplitude_shock_cold} shows the shock wave amplitude. It decays considerably due to dissipation of the wave energy in the shock front. For comparison we also plot variation of the wave amplitude calculated with the assumption that there was no dissipation in the shock front. These plots demonstrate that a significant part of the shock wave energy is spent on plasma heating.

\begin{figure} % {7}
  \centerline{\includegraphics[width=0.5\textwidth]{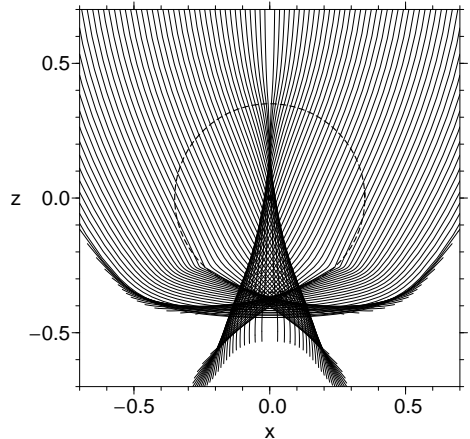} }
\caption{The propagation of a shock wave near a null point in a warm
plasma. The dashed line denotes the location of the $V_A=c$~layer,
the solid ones are the rays. At the initial moment, the plane wave
front was located at $z_0=1$ and $-1 \leq x \leq1$. The axes
are shown in the normalised units.}
  \label{F-prop_pattern_shock_warm}
\end{figure}

We have also considered the propagation of a weak shock
wave in a warm plasma. In this case both nonlinearity of the wave
and the non-zero sound speed in a plasma work in the same way. The
shock wave propagates through the null point and a caustic is
formed. The ray pattern is shown in Figure
\ref{F-prop_pattern_shock_warm}. The caustic is quite similar to
that in Figure \ref{F-prop_pattern_lin_warm} for the linear warm
plasma case, however it is shifted in the direction of the wave
incidence. An effect of the non-zero sound speed appears to
predominate over that of the wave nonlinearity. This is not unusual
if we keep in mind that the nonlinear geometrical acoustics method
should be applied only for weak shock waves. We also note
that two caustic cusps are located at the $V_A=c$~layer.

\section{Discussion and Conclusions}
\label{S-conclusion}

We have applied the nonlinear geometrical acoustics method to model
the propagation of a fast-mode MHD wave near a 2D magnetic null
point. We have considered both a linear wave and a weak shock and calculated their behaviour in cold and warm plasmas. We have also
calculated the wave amplitude, using the ray approximation and the
laws of solitary shock wave damping.

The nonlinear ray method allows one to trace the wave front and calculate the jumps of MHD quantities at the front only, but it does not give information on the downstream plasma flow evolution, since irreversible processes occur behind the shock front. In the propagation model used taking into account the refraction effect (\textit{i.e.} the wrapping effect), the wave front can encounter plasma ``upstream'' that at an earlier time was described as ``downstream''. In this sense, the presented nonlinear ray equations are valid for weak shock waves, if there are no subsequent shock front intersections. Nevertheless, to investigate plasma heating we calculate the caustic pattern, neglecting the modified downstream plasma, which is correct for weak shock waves. We note again that since subsequent evolution of the downstream plasma flow cannot be analysed with the method used, we do not discuss the transformation of a null point into a current sheet under the shock wave effect.

For a linear wave approaching the null point in a
cold plasma, the analytical law of the amplitude growth has been
found. The wave amplitude increases exponentially and therefore the
initially linear wave transforms into a shock, which has
encouraged us to investigate the case of shock wave propagation. This analytical law expressed by Equation~(\ref{E-wave_amp_expr_u}) (and Equation~(\ref{E-wave_amp_expr_magnitude})) is consistent with the analytical solution that was obtained for the problem of the incidence of a cylindrically symmetric fast-mode wave on a 2D null point (\opencite{Syrovatskii66}; \opencite{Craig91}; \opencite{Longcope07}). Indeed, in their solution the radial component of the plasma velocity in the wave grows as $1/r$ with the distance $r$ to the null point. In the cylindrically symmetric case all the rays are straight and directed to the null point. So we have the following ray equation: $dr/dt=-V_A=-r$, and hence we obtain: $r \propto e^{-t}$, \textit{i.e.}, the wave amplitude grows exponentially, like that in Equation~(\ref{E-wave_amp_expr_u}). The additional time dependence in Equation~(\ref{E-wave_amp_expr_u}) is due to the plain (not cylindrical) initial wave front and represents the fact that all the rays are different and depend on initial coordinate $x_0$.

A complex caustic is formed around the magnetic null point. The wave
amplitude tends to infinity at the caustic, resulting in plasma
heating. The heating is distributed in space and occurs mainly out
of the $V_A=c$~layer if the initial wave front is rather extensive.
However, the first heating associated with a central part of the
wave occurs within the $V_A=c$~layer. On the contrary, in the
well-known case of a linear wave in a cold plasma, plasma heating
takes place in a small vicinity of the null point. Since real null
points in the solar corona appear to capture only a (central) part
of the wave, while its outer part propagates away, the size of the
$V_A=c$ layer may be chosen as the spatial scale of plasma heating.

Nonlinearity of a wave is crucial for its propagation. A shock wave
is able to pass through a null point even in a cold plasma. The
growth of the shock wave amplitude is restricted, since shock waves
decay significantly due to the nonlinear energy dissipation in the
shock front. Therefore along with the heating at a caustic, a
substantial heating of plasma occurs in the neighbourhood of the null
point. A significant part of the shock wave energy transforms into
heat there.

Drawing the caustic pattern, we have explained the complicated
surface of the wave front. A break of the wave front occurs at the
caustic, thus triangles of the wave front arise when the front
passes through a caustic cusp. The vertices of the triangles travel
along the caustic lines.

To summarise we discuss a chance to estimate the magnitude of plasma
heating associated with propagation of waves near a magnetic null
point. Conversion of the wave energy into heat occurs due to
dissipative processes in a plasma when the gradients of MHD
quantities increase significantly. In a linear wave the gradients
grow owing to the wave extent decrease as well as the amplitude
growth as the wave approaches a null point. To estimate the value of
the energy converted into heat in this case, one should consider
non-ideal magnetohydrodynamics. In turn, in a shock wave the
gradients of MHD quantities also originate at the wave front due to
the nonlinear steepening of the wave profile. Here, the value of the
energy converted is taken into account by the laws of shock wave
damping. Since the sound speed in the real solar coronal plasma does
not drop to zero, the wave extent can remain rather large, and so 
viscosity, thermal conductivity and finite plasma conductivity
do not result in energy conversion. In such a case considerable plasma
heating will occur due to the wave energy dissipation in the shock
front, with the nonlinear heating predominating over heating at
caustics. So, evaluating the difference between the shock wave
amplitude and the amplitude calculated without dissipation in the
discontinuity (as in Figure~\ref{F-amplitude_shock_cold}), one can
provide some rough estimates about the plasma heating magnitude.

%%%%%%%%%%%%%%%%%%%%%%%%%%%%%%%%%%%%%%%%%%%%%%%%%%%%%%%%%%%%%%%%%%%%%%%%%%%
\appendix

\section{Ray equations}
\label{S-appendix1}

The sets of ray equations have the following form in Cartesian
coordinates:
%%%%%%%%%%%%%%%% nonlin ray equations system %%%%%%%%%%%%%%%%%%%%%%%%%%%%%%%%%
\begin{eqnarray}
 \frac{dx}{dt}&=&\left( a+N \frac{\kappa U_{sh}}{2}\right) \frac{k_{x}}{k}
 +k\frac{\partial a}{\partial k_{x}},\quad\quad\quad \frac{dk_{x}}{dt}=-\frac{\partial a}
 {\partial x}k, \nonumber \\
 \frac{dz}{dt}&=&\left( a+N \frac{\kappa U_{sh}}{2}\right) \frac{k_{z}}{k}
 +k\frac{\partial a}{\partial k_{z}}, \quad\quad\quad  \frac{dk_{z}}{dt}=-
 \frac{\partial a}{\partial z}k,
\label{E-ray_eq_system}
\end{eqnarray}
%%%%%%%%%%%%%%%%%%%%%%%%%%%%%%%%%%%%%%%%%
where $k_x, k_z$ and $k$ are the components of the wave
front normal $\mathbf{k}$ and its magnitude respectively,
and $N$ the numerical factor switching the linear and nonlinear
regimes of the wave propagation. We should take $N=0$ for studying a
linear wave and $N=1$ for a weak shock one.

The adjoint set consists of differential equations for the
derivatives $\partial x/\partial \eta$, $\partial z/\partial \eta$,
$\partial k_x/\partial \eta$, $\partial k_z/\partial \eta$ and is
derived by differentiating Equations (\ref{E-ray_eq_system})(with
$N=0$) with respect to the ray coordinate $\eta$:
%
%
%%%%%%%%%%%%%%%%%%%%% equations system for calc amp %%%%%%%%%%%%%%%%%%%%%%%%%%%
\begin{eqnarray}
\label{E-adj_system}
\frac{d}{dt}\left( \frac{\partial x}{\partial \eta }\right)
=\frac{k_{x}}{k}\frac{\partial a}{\partial \eta }+%
\frac{a}{k}\frac{\partial k_{x}}{\partial \eta }-
\frac{a k_{x}}{k^{2}}\frac{%
\partial k}{\partial \eta }+
\frac{\partial a}{\partial k_{x}}\frac{\partial k%
}{\partial \eta }+k\frac{\partial }{\partial \eta }\left(
\frac{\partial
a}{\partial k_{x}}\right), \nonumber \\
\frac{d}{dt}\left( \frac{\partial z}{\partial \eta }\right)
=\frac{k_{z}}{k}\frac{\partial a}{\partial \eta }+%
\frac{a}{k}\frac{\partial k_{z}}{\partial \eta }-
\frac{a k_{z}}{k^{2}}\frac{%
\partial k}{\partial \eta }+
\frac{\partial a}{\partial k_{z}}\frac{\partial k%
}{\partial \eta }+k\frac{\partial }{\partial \eta }\left(
\frac{\partial
a}{\partial k_{z}}\right), \nonumber \\
\frac{d}{dt}\left( \frac{\partial k_x}{\partial \eta }\right)
=-\frac{\partial k}{\partial \eta}\frac{\partial a}{\partial x}- k
\frac{\partial}{\partial \eta} \left( \frac{\partial a}{\partial x
}\right), \\
\frac{d}{dt}\left( \frac{\partial k_z}{\partial \eta }\right)
=-\frac{\partial k}{\partial \eta}\frac{\partial a}{\partial z}- k
\frac{\partial}{\partial \eta} \left( \frac{\partial a}{\partial z
}\right), \nonumber
\end{eqnarray}
where
\begin{eqnarray*}
\frac{\partial a}{\partial \eta }=\sum\limits_{\alpha }\left(
\frac{\partial a}{\partial r_{\alpha }}\frac{\partial r_{\alpha
}}{\partial \eta }+\frac{
\partial a}{\partial k_{\alpha }}\frac{\partial k_{\alpha }}{\partial \eta }
\right), \quad\frac{\partial k}{\partial
\eta}=\frac{1}{k}\sum\limits_{\alpha }k_{\alpha }\frac{\partial
k_{\alpha }}
{\partial \eta},  \\
\frac{\partial }{\partial \eta }\left( \frac{\partial a}{\partial k_{\beta }%
}\right) =\sum\limits_{\alpha }\left( \frac{\partial }{\partial r_{\alpha }}%
\left( \frac{\partial a}{\partial k_{\beta }}\right) \frac{\partial
r_{\alpha }}{\partial \eta }+\frac{\partial }{\partial k_{\alpha
}}\left(
\frac{\partial a}{\partial k_{\beta }}\right) \frac{\partial k_{\alpha }}{%
\partial \eta }\right),  \\
\frac{\partial }{\partial \eta }\left( \frac{\partial a}{\partial r_{\beta }%
}\right) =\sum\limits_{\alpha }\left( \frac{\partial }{\partial r_{\alpha }}%
\left( \frac{\partial a}{\partial r_{\beta }}\right) \frac{\partial
r_{\alpha }}{\partial \eta }+\frac{\partial }{\partial k_{\alpha
}}\left(
\frac{\partial a}{\partial r_{\beta }}\right) \frac{\partial k_{\alpha }}{%
\partial \eta }\right),  \\
r_{\alpha}=x,z; \; \; k_{\alpha}=k_x,k_z. \qquad \qquad \qquad
\qquad
\end{eqnarray*}

\begin{figure} % {8}
  \centerline{\includegraphics[width=0.7\textwidth]{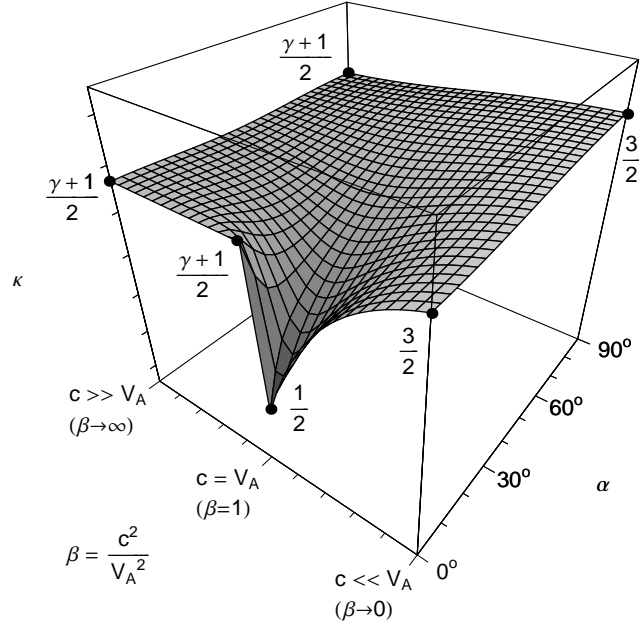} }
\caption{The values of coefficient $\kappa$. The points at the
surface denote the relevant values corresponding to some particular
cases of propagation. There is a jump of $\kappa$ at $\alpha=0$ and
$\beta=1$.}
  \label{F-kappa_surface}
\end{figure}

The numerical coefficient $\kappa$ depends on the plasma beta as
well as the angle $\alpha$ between the wave front normal
and the magnetic field as
$$\kappa =\frac{(1+\beta +Q_{1})(\beta (\gamma -2)+3Q_{1})-2\beta \cos
^{2}\alpha (\gamma -1-\beta +Q_{1})}{2Q_{1}(1+\beta +Q_{1})},$$
where $Q_{1} =\sqrt{(1+\beta )^{2} -4\beta \cos ^{2} \alpha }$, $\beta=c^2/V_A^2$, and $\gamma=5/3$ is the adiabatic index. Values of $\kappa$ are restricted by the limits $1/2\leq \kappa \leq 3/2$ and are shown in Figure \ref{F-kappa_surface}. When the wave crosses  the $V_A=c$~layer along the magnetic field ($\alpha=0$), $\kappa$ has a jump.

Set of ray equations~(\ref{E-ray_eq_system}) is not closed in the
nonlinear case since it includes the wave amplitude $U_{sh}$. It can
be calculated with the laws of solitary shock wave damping
(\opencite{Uralov82}), which are derived by using values of the
amplitude and duration of a simple wave from which the shock forms:
\begin{equation}
U_{sh}=u_{1}\left( 1+\frac{\tau _{1}}{T_0}\right)^{-1/2}, \qquad \frac{d\tau _{1}}{dt}=\frac{\kappa u_{1}}{a},
\label{E-damping_laws}
\end{equation}
where $\tau_{1}$ is the duration increment of the simple wave with
amplitude $u_{1}$; $T_0$ is the initial duration of the shock wave.
The value of $u_{1}$ can be found from an equation similar to
Equation (\ref{E-dS_wave_amp_variation}).

Solving sets of ordinary differential equations (\ref{E-ray_eq_system}),~(\ref{E-adj_system})~and~(\ref{E-damping_laws}) numerically we are able to model the propagation of fast-mode MHD waves.

\section{Derivation of nonlinear ray equations}
\label{S-appendix2}

The sets of ray equations can be obtained with a different method
which does not use the eikonal expansion of the MHD equations. In
fact, this method is different only in form and it is more
convenient for analysing nonharmonic waves which we deal with in
this paper. We apply this method for deriving the set of nonlinear
ray equations.

The moving surface of the wave front can be described by the equation $\Phi(\mathbf{r},t)=0$, or in the differential form $d \Phi = 0$. Let a small surface element displace by $d \mathbf{r}$ in time $d t$. Then we have:
\begin{equation}
\frac{\partial \Phi }{\partial t} + \left ( \mathrm{grad}\Phi \cdot \frac{d \mathbf{r}}{d t} \right )= 0,
\label{E-surface_eq}
\end{equation}
where $d \mathbf{r} / d t = \mathbf{q}$ is the velocity of the front element and $\mathrm{grad}\Phi$ the wave front normal by definition. The scalar product can be rewritten as $q_n \left\vert \mathrm{grad}\Phi \right\vert$, with the $q_n$ denoting the normal speed of the wave front. So we obtain a partial differential equation.

If we consider a linear fast-mode MHD wave, $q_n = a$, and we have
the same ``eikonal'' equation and the same set of ray equations as
presented in Section~\ref{S-linear wave-cold plasma} and Appendix
\ref{S-appendix1} (the case of $N=0$), respectively. However, we are
interested in the weak shock wave propagation. The normal speed of
the weak shock front can be derived from the jump conditions across
a MHD discontinuity, and it is equal to $a + \kappa U_{sh}/2$ in the
nonlinear acoustics approximation. Substituting it into
Equation~(\ref{E-surface_eq}) and solving with the method of
characteristics, we obtain the set of ray equations, which is
similar to the linear one and in which substitution $a \rightarrow a
+ \kappa U_{sh}/2$ has been performed. In these new ray equations
the additional terms $\partial U_{sh}/\partial k_\alpha$ and
$\partial U_{sh}/\partial r_\alpha$ appear. We neglect them, keeping
only quantity $U_{sh}$ itself and omitting its derivatives,
exploiting the nonlinear geometrical acoustics approximation. Thus
we obtain the set of nonlinear ray equations (\ref{E-ray_eq_system})
(with $N=1$) presented in Appendix~\ref{S-appendix1}.

%%%%%%%%%%%%%%%%%%%%%%%%%%%%%%%%%%%%%%%%%%%%%%%%%%%%%%%%%%%%%%%%%%%%%%%%%%%

\begin{acks}

We thank the anonymous referee for helpful comments and valuable
suggestions. We also thank the editors for careful reading of the
manuscript and useful comments. A.A. is very grateful to the
scientific organising committee of the ESPM--13 for financial
support.

The research was supported by the Russian Foundation of Basic
Research (Grant No. 12-02-00037) and a Marie Curie International Research
Staff Exchange Scheme Fellowship within the 7th European Community
Framework Programme as well as Siberian Branch of the Russian
Academy of Sciences (Lavrentyev Grant 2010--2011). It was also supported by the Ministry of Education and Science of the Russian Federation (State Contracts 16.518.11.7065 and 02.740.11.0576).

\end{acks}

%%% BIBLIOGRAPHY %%%%%%%%%%%%%%%%%%%%%%%%%%%%%%%%%%%%%%%%%%%%%%%%%%%%%%%%%%%

\end{article}
\end{document}